\documentclass[12pt,a4paper]{article}
\usepackage{graphics}
\usepackage{latexsym}

\begin{document}
\newcommand{\ket}[1] {\mbox{$ \vert #1 \rangle $}}
\newcommand{\bra}[1] {\mbox{$ \langle #1 \vert $}}
\def\M{{\mbox{M}}}
\def\Mb{\bar{\mbox{\bf M}}}
\def\mb{\bar{\mbox{\bf m}}}
\def\m{{\mbox{m}}}
\def\mub{\bar \mu}
\def\Ar {{\mbox{\bf A}}(k \vert k^{\prime }, k^{{\prime }{\prime }})}
\def\Arm{{\mbox{\bf A}}(-k|-k^{\prime },-k^{{\prime }{\prime }})}
\def\Br {{\mbox{\bf B}}(k^\prime \vert k, -k^{{\prime }{\prime }})}
\def\Cr {{\mbox{\bf C}}(k^{{\prime }{\prime }} \vert k, -k^{\prime })}
\def\Vr {{\mbox{\bf V}}(\vert -k,k^{\prime },k^{{\prime }{\prime }})}
\def\Vrm{{\mbox{\bf V}}(k,k^{\prime },k^{{\prime }{\prime }})}
\def\fin{\end{document}}
\def\lrD{\mathrel{{\cal D}\kern-1.em\raise1.75ex\hbox{$\leftrightarrow$}}}
\def\lr #1{\mathrel{#1\kern-1.25em\raise1.75ex\hbox{$\leftrightarrow$}}}
\def\lrpartial{\mathrel
{\partial\kern-.75em\raise1.75ex\hbox{$\leftrightarrow$}}}
\newcommand{\be} {\begin{equation}}
\newcommand{\ee} {\end{equation}}
\newcommand{\ba} {\begin{eqnarray}}
\newcommand{\ea} {\end{eqnarray}}
\begin{titlepage}
\begin{flushright}
UMH-MG-97/01\\
TAUP 2428-97
\end{flushright}
\vskip 1.cm
\begin{center}
{\LARGE\bf Interacting Charged Particles in an 
Electric Field and the
Unruh Effect}\\
\vskip 1.cm
{Cl. Gabriel\footnote{Aspirant du FNRS} \footnote{E-mail
gabriel@sun1.umh.ac.be}, 
Ph. Spindel\footnote{E-mail spindel@sun1.umh.ac.be}},
\\
\vskip 0.5 cm
 {\em M\'ecanique et Gravitation}\\
 {\em Universit\'e de Mons-Hainaut}, \\
{\em 15, avenue Maistriau, B-7000 Mons, Belgium}
\vskip 0.75 cm
 {S. Massar\footnote{E-mail massar@post.tau.ac.il}}\\
\vskip 0.5 cm
{\em Raymond and Beverly Sackler Faculty of Exact Sciences},\\
{\em School of Physics and Astronomy},\\
{\em Tel-Aviv University, Tel-Aviv 69978, Israel}
\vskip 0.75 cm
 {R. Parentani\footnote{E-mail parenta@celfi.phys.univ-tours.fr}}\\
\vskip 0.5 cm
{\em Laboratoire de Math\'ematiques et  Physique th\'eorique},\\
{\em  CNRS-UPRES A 6083, 
Facult\' e des Sciences},
\\{\em Universit\'e de Tours, 37200 Tours, France}
\vskip 1 cm
\end{center}
\pagebreak
\begin{abstract}

We compute the transition amplitudes between 
charged particles of mass $M$ and $m$
accelerated by a constant electric field and interacting by
the exchange of quanta of a third field. We work in second
quantization in order to take into account both recoil effects
induced by transitions and the vacuum instability of 
the charged fields. In spite of both effects, when the
exchanged particle is neutral, the equilibrium ratio of the populations 
is simply $\exp(\pi (M^2 - m^2)/eE)$. 
Thus, in the limit
$(M-m)/M \to 0$, one recovers Unruh's result characterized by the
temperature $a/2\pi$ where $a$ is the acceleration.
When the exchanged particle is charged, its vacuum instability
prevents a simple description of the equilibrium state. 
However, in the limit
wherein the charge of the exchanged particle tends to zero,
the equilibrium distribution is once more 
Boltzmanian, but characterized not only by a temperature but also by the 
electric potential felt by the exchanged particle. This work therefore
confirms that thermodynamics in the presence of 
horizons does not rely on a semi-classical treatment. 
The relationship with thermodymanics of charged black holes
is stressed.

\end{abstract} 
\end{titlepage}

\section{Introduction}

Shortly after Hawking's seminal discovery of black hole 
radiation\cite{Hawk},
Unruh\cite{Unr} showed that it possesses a flat space
analogue, namely that a uniformly accelerated detector perceives
Minkowski vacuum to be thermaly populated at temperature $T_U =
a/2\pi$. In Unruh's original work, only the detector's internal states
where treated quantum mechanically. Its position was treated classically
and thus insensitive to the transitions. 
This is clearly an approximation since it violates momentum conservation.
In order to analyze the validity of this approximation
and to re-establish  momentum conservation,
one must treat
the detector position quantum 
mechanically\cite{rec,rez,ScUn}. 
This enlargement of the quantum dynamics to a variable
 formally external to the dynamics provides
new insights about the Unruh process.
Moreover, it may also be used as a guide to other
problems dealing with particle
creation in the presence of horizons since, in those cases,
 similar approximation schemes are used and should be 
abandoned in other to address the question
of the quantum back reaction.
The new insights concerning the Unruh effect 
which have been obtained are:
\begin{enumerate}
\item The detector can be described by a delocalised
wave function, whereupon the classical geometric notion of a horizon
no longer exists. (Of course, one may approximatively recover
the concept of horizon upon building well localized wave packets.)
Nevertheless thermal rates for transitions of the
detector still obtain thereby confirming that thermodynamical relations
still govern the physics when one goes beyond the semiclassical 
treatment. 
\item Each time the detector makes a transition, it
recoils both in momentum and in energy 
in such a way that
the total instantaneous Minkowski momentum and energy is conserved in the
sense explained in \cite{rec}, pages 245-246.
\item These recoils give rise to a decoherence of 
the detector-radiation system. This in turn implies that the detector
emits a steady flux of radiation, contrary to the situation where
the
detector's position is treated classically  \cite{grove,rsG,mpb}. 
\item The Unruh effect and the
Schwinger process\cite{schw1,schw} 
of pair creation in an electric field are ``in
equilibrium''\cite{bps}. 
Moreover the area of the acceleration horizon plays the r\^ole of an entropy
in delivering 
the equilibrium population ratios.
\end{enumerate}
To understand the dynamical origin of these equilibrium notions is one of 
 our main concerns. Thus,
we first present the content of point 4 in some detail.
In \cite{ScUn}, the accelerated detector is modeled by a ``two level
ion'' propagating in constant electric field $E$.
It has charge $Q$ and its two
levels  have rest mass $M$ and $m$. It therefore uniformly accelerates
with 
acceleration $Q E/ M$ or $Q E/ m$ according to its mass. 
The ion can make transitions between
its two levels by emitting or absorbing a massless quantum. Thus it
behaves like an accelerated particle detector with mass gap $\Delta M = M -m$. 
Moreover, when describing the ion's states by field operators, i.e. 
by working in second quantization,
the electric field leads to vacuum instability through
 pair creation of ions anti-ions.
The mean numbers of created pairs 
of mass $m$ and $M$ are~:
\be
N_{m} = e^{- \pi m^2 /QE}\quad , \quad 
N_{M} = e^{- \pi M^2 /QE}\qquad .
\label{Sch}
\ee
A priori independently of these creation processes, an ion will make
transitions 
from its excited to its ground state at a rate $R_{M\to m}$, 
or from its ground to its excited state at a rate $R_{m \to M}$. 
As suggested in \cite{bps} and confirmed in \cite{ScUn,MeLic},
these transitions are intimately related to the creation processes
since the ratio of their rates is 
given by
\be
{ R_{m \to M} \over R_{M\to m}}
= {N_{M} \over N_{m}} = e^{- \pi (M^2 -m^2) /QE} \qquad .
\label{Rates2}
\ee
Thus 
we can say that the two
processes are in equilibrium since they determine the same distribution. 
We emphasize that  eq. (\ref{Rates2}) is exact in the sense that it takes
into account
all effects due to the
finite mass of the detector, 
i.e. recoil effects, and the finite probability to create
pairs of detectors.

Upon taking the limit $M,m \to \infty$, with $\bar a =  2 Q E / (M + m )$ and
$\Delta M $ constant, recoil effects and pair creation amplitudes vanish.
Therefore, one expects to recover Unruh's result 
which gives the equilibrium probabilities
of an accelerated detector of ``infinite'' mass and of given acceleration
$\bar a$.
Indeed, upon taking the above limit, eq. (\ref{Rates2}) becomes
\be
{ R_{m \to M} \over R_{M\to m} }
= e^{ -  \pi \Delta M (M+m) / Q E} = e^{- 2 \pi \Delta M / \bar a}\qquad .
\label{Rates}
\ee

The interest of eq. (\ref{Rates2}) is further enhanced when one recalls
that the probability for pair creation due to the 
Schwinger process can be expressed in terms of the
change of the area of the acceleration horizon\cite{HHor,Inst}
\be
N_{m} = e^{- \Delta A_H (m,Q)}
\label{a4}
\ee
where $\Delta A_H (m,Q) = A_H^0 - A_H(m,q) $ 
is the area of the acceleration
horizon in flat space minus the area 
if a pair of particles of mass $m$ and charge $e$ is emitted\footnote{
More specifically it is the change of the area of the horizon in the
euclidean continuation of space time, see the cited references for
additional details}.
Thus eq. (\ref{Rates2}) can be rewritten as
\be
{ R_{m \to M} \over R_{M\to m}}
= e^{- (\Delta A_H (M,Q)- \Delta A_H(m,Q))}=  
e^{- \Delta A_H(\Delta M) }
\label{Rates3}
\ee
where in the last equality we have written 
$\Delta A_H(\Delta M) = A_H(m,Q) - A_H(M,Q)$ as the difference
of the horizon area between the initial and final states.
The area of the acceleration horizon therefore behaves like 
an entropy 
in delivering the equilibrium population ratios.

The aim of the present paper is to extend these notions
to the situation wherein the field $\Phi$ with which the 
detector interacts is charged and therefore feels also the 
constant $E$-field.
Our main results 
are:
\begin{enumerate}
\item
The transition rates and the equilibrium distribution are no
longer described by Boltzman ratios. Hence the connection with
thermodynamics no longer obtains.
Similarly the equilibrium relation, eq. (\ref{Rates2}) between
 Schwinger and radiative processes is broken.
\item
However in the case of weakly charged exchanged particles,
i.e. for
$\alpha /Q <\!\!< 1$ where $\alpha$ is the charge of
the field $\Phi$, an extended thermodynamical relation
is obtained. Indeed, in that case, eq. (\ref{Rates}) becomes
\end{enumerate}
\be
{ R_{m \to M} \over R_{M\to m}}
= e^{-{ 2\pi \over \bar a} ( \Delta M - \alpha { E \over 2 \bar a})
+ O(\alpha^2 /Q^2)} \qquad ,
\label{Rates5}
\ee
In a forthcoming article
we shall confirm these results by adopting 
the more kinematical point of view of quantizing 
$\Phi$
in Rindler coordinates\cite{Charge}.

The thermodynamical relation eq. (\ref{Rates5})
 shows that, in addition to the Unruh temperature $T_U = \bar a / 2
\pi$, there is now an electric potential $( = E/
2 \bar a)$ which modifies the equilibrium.
This
is strictly analogous to the ratio of the rates
for charged particles to be emitted or absorbed by a charged black hole.
In Hawking's derivation, this ratio is expressed in terms of the 
Bogoljubov coefficients $\gamma_{\omega}, \beta_{\omega}$
characterizing the mixture of {\it in} and {\it out} modes
of the $\Phi$ field.
However, to make clearer the contact with eq. (\ref{Rates5}),
we can express Hawking's  result in terms of the rate
${\cal{R}}_{M \to M-\omega}$ to jump from a
black hole of mass $M$ and charge $Q$ to a black hole characterized by 
 $M-\omega$ and $Q -\alpha$ and the rate of the inverse process
${\cal{R}}_{M-\omega \to M}$. 
Indeed, Hawking's result can be expressed as  
\be 
\vert {\beta_{\omega} \over \gamma_{\omega} }\vert^2 
= e^{\beta_H (\omega - \alpha \phi )} = {{\cal{R}}_{M - \omega \to M} \over
{\cal{R}}_{M\to M - \omega}}
\label{Rates51}
\ee 
where $\beta_H$ is the inverse
Hawking temperature, $\omega$ is the energy of the 
quantum measured at spatial infinity and $\phi$ is the difference of
electric potential between the
horizon and infinity.
In eq. (\ref{Rates5}),
the equivalent of $\phi$ is, ${ E /2 \bar a}$, the difference of
electric potential between the
horizon and the accelerated trajectory where the charged quantum is emitted
(absorbed).
Indeed, in Rindler coordinates $\tau$, $\rho$, the (static) potential is
\be
A_\tau (\rho)= E \bar a \rho^2 /2
\ee
and the particle is localized at $\rho = 1/ \bar a$.

We wish to emphasize that in both cases the thermodynamical 
canonical relations, eqs. (\ref{Rates5}, \ref{Rates51}), can be a posteriori 
related to changes in horizon's area.
Indeed, in both cases, the logarithm of these ratios is equal to the a quarter 
of the horizon area change when linearized in the energy and charge
differences, i.e. in the energy and charge carried by the quantum of the
$\Phi$ field. In the black hole case, this is a reexpression of the first
law of black hole thermodynamics. In the present case, 
since eq. (\ref{Rates5}) deals with the change in the accelerated horizon, 
\be
{\Delta A_H (\Delta M, \alpha) \over 4} =_{linearized} { 2\pi \over \bar a}
( \Delta M - \alpha { E \over 2 \bar a})
\label{accfl}
\ee
should be considered as the (linearized)
 first law of accelerated-horizon thermodynamics.

However, in contradistinction to the
treatment used to derive eq. (\ref{Rates51}), we shall not make use of any
background field approximation to obtain eq. (\ref{Rates5}). Instead,
we shall approximate the exact result by taking variations limited 
to first order in $\alpha$ and $\Delta M$, see Section 4.
Therefore, we can analyze the finite difference and not only
the first order changes delivering the above canonical concepts through
differentiation.
Upon neglecting the vacuum instability of the 
exchanged $\Phi$ field, we obtain that the logarithm of the
ratio of the transition rates is given by the following finite difference
\be
{\Delta A_H \over 4} = { \pi } ( {M^2 \over QE }- {m^2 \over qE})
\label{accfl2}
\ee
where $q= Q - \alpha$ is the ``residual'' charge of the ground state of
mass $m$.
Since eq. (\ref{accfl2}) agrees with the Schwinger process, see eq. (\ref{Sch}),
by virtue of eq. (\ref{a4}), this finite difference is equal to the
 finite change in area, as in eq. (\ref{Rates3}).
This confirms that a quarter of the area not only delivers canonical
distributions and thermodynamics but truly determines quantum processes as
in statistical mechanics. 

To conclude let us recall once more 
the origin of the agreement
of eq. (\ref{Rates}) with Unruh's result (or the agreement of eq.
(\ref{a4}) with
eq. (\ref{Rates3})). Eq. (\ref{Sch}) is
derived
in an enlarged  quantum setting wherein the trajectory of the heavy ion is 
quantum mechanically treated whereas Unruh's derivation is based
on a background field approximation since the trajectory is classically
determined.
The agreement of the ratio of the transition rates evaluated from both 
treatments arises when the following procedure is applied to transition
amplitudes evaluated in the more quantum framework.
Upon working with WKB waves and
performing first order expansions in $\Delta M/M$ and in the momentum transfer,
these transition amplitudes coincide with the corresponding amplitudes
 evaluated at the background field approximation, see \cite{rec} for the
details. 
What guarantees this agreement is that first order expansions of the 
WKB phases are controlled by Hamilton-Jacobi equations.
The same relation will therefore hold when one considers 
gravity. (This has been explicitly verified in mini-superspace in
\cite{wdwpt}.)
To first order in the matter energy change, transition amplitudes computed
in quantum gravity with WKB waves are equal to the corresponding 
amplitudes evaluated from quantum field theory in a given 
classical geometry. Only second order changes, i.e. the non-linear response
of gravity, involve the Planck mass. 

The present article is organized as follows. Section {\bf 1} is this
introduction. Section {\bf 2} is devoted to recalling the quantization of a
charged field in an electric field and the Schwinger process. In Section {\bf 3}
we introduce our detector model, and consider the case when the
detector interacts with a massive, but neutral, field. The techniques
developed in this section are then used in Section {\bf 4} to analyze the
more complicated case of interactions with a charged field. Appendix {\bf 1} 
is devoted to the analytical evaluation of the transition amplitudes and
Appendix {\bf 2}
deals with the limit $\alpha \mapsto 0$ of the transition amplitudes.

\section{Charged Particles in an Electric Field}
The aim of this section is to review the quantization of a massive
charged scalar field in an external electric field $E$. For the reader
interested in a more complete treatment we refer to \cite{GO} and
references therein.
Classically, the equation of motion of a relativistic charged particle of 
mass $M$ and charge $Q$ in  an electric field is:
\begin{eqnarray}
M\frac{d^2x^\mu }{d\tau ^2}=-QE \ \varepsilon^{\mu \nu }\frac{dx_\nu }{d\tau } 
\end{eqnarray}
Its trajectory is a hyperbola with parametric equations  given by:
\begin{eqnarray}
t+\frac k{QE}=\frac 1A\sinh A(s-s_0)\quad,\\
x-x_0=\frac 1A\cosh A(s-s_0)\qquad . \label{traj}
\end{eqnarray}
Here $A=QE/M$ is the classical acceleration of the particle 
The time coordinate of the turning point of the
hyperbola is $t_0=-k/QE$ where $k=(M\dot x +A_x)$ is the conserved momentum
canonically conjugate to the variable $x$ in the gauge $A_t=0$
and $A_x=-Et$.

The corresponding Klein-Gordon equation for the field is:
\begin{eqnarray} \left[
\Box 
+ M^2\right] \psi _M(t,x)\equiv \left[ (-i\partial
_x+QEt)^2+\partial _t^2+M^2\right] \psi _M(t,x)=0 \label{field} 
\end{eqnarray}
The general solution of this equation can be written as a
superposition of modes:
\begin{eqnarray}
\psi _{M,k}(t,x)=\frac{e^{ikx}}{\sqrt{2\pi }}\chi _{M,k}(t)\quad,
\end{eqnarray}
and their complex conjugates. The functions $\chi_{M,k}(t)$ obey the equation:
\begin{eqnarray}
\left[\partial _t^2 +M^2+(k+QEt)^2\right] \chi _{M,k}(t)=0\quad,
\end{eqnarray}
Because of the $t$ dependence of
this potential there will be some backscattering. To interpret this in
the context of the Klein-Gordon equation, we note that the current
operator is $J_0 = -iQ \lrpartial_t$. Thus the backscattering
corresponds to mixing of positive and negative charge. In a second
quantized context this corresponds to pair creation \cite{bps,GO}.
To describe them
we introduce a set of solutions with only positive or negative
charge for $t\to - \infty$ (in-modes)
\begin{eqnarray}
\psi _{M,k}^{p\,in}(t,x)&=&\frac{e^{ikx}}{\sqrt{2\pi }}\frac{e^{-\frac \pi
4 \epsilon
_M}}{(2QE)^{1/4}}D_{i\epsilon _M-1/2}\left[ e^{3i\pi /4}(t+\frac
k{QE})\sqrt{2QE}\right]\nonumber\\
&=& \frac{e^{ikx}}{\sqrt{2\pi }} {\cal{D}}_{\epsilon _M}[\lambda]
\label{modepin}\\ 
\psi _{M,-k}^{a\,in}(t,x)&=&
\frac{e^{-ikx}}{\sqrt{2\pi }} {\cal{D}}_{\epsilon _M}[\lambda]
\label{modeain}
\end{eqnarray}
and a set of solutions with only positive or negative
charge for $t\to + \infty$ (out-modes)
\begin{eqnarray}\psi _{M,k}^{p\,out}(t,x)&=&
 \left[ \psi _{M,-k}^{p\,in}(-t,x) \right]^*\nonumber\\
&=&\frac{e^{ikx}}{\sqrt{2\pi }} {\cal{D}}_{\epsilon _M}^*[-\lambda]
\label{modepout}\\ 
\psi _{M,-k}^{a\,out}(t,x)&=& \left[ \psi _{M,-k}^{a\,in}(-t,x)
\right]^*
\nonumber\\
&=&\frac{e^{-ikx}}{\sqrt{2\pi }} {\cal{D}}_{\epsilon _M}^*[-\lambda]
\label{modeaout} \end{eqnarray} where $\epsilon_M=M^2/(2QE)$
and where the superscripts $p$ and $a$ refer respectively to particle and
anti-particle wave functions. 
We have introduced a synthetic notation for the 
parabolic cylinder
functions and their argument
\begin{eqnarray}
{\cal{D}}_{\epsilon _M}[\lambda]&=&
\frac{e^{-\frac \pi 4 \epsilon_M}}{(2QE)^{1/4}} 
D_{i\epsilon _M-1/2}\left[ e^{3i\pi
/4}(t+\frac k{QE})\sqrt{2QE}\right]
 \nonumber\\
 \lambda &=& (t+\frac k{QE})\label{Par}
\end{eqnarray}
We also note that the  parabolic cylinder
function have the following integral representation\cite{GO}~: 
\begin{equation}
{\cal{D}}_{\epsilon _M}[\lambda]=
\frac{e^{-\frac \pi 2 \epsilon_M}e^{-i\pi /8}}{(2QE)^{1/4}
\Gamma (\frac 12 -i\epsilon _M)} 
e^{+\frac i2 \lambda^2 QE}
\int_0^\infty dv\ e^{-i\lambda \sqrt{2QE}v+i\frac{v^2}2}
v^{-i\epsilon _M -\frac 12} \label{DIrep}
\end{equation} 
where the integration parameter $v$ is classically related to
$t$ and its conjugate momentum $p_t$ by\cite{BaVo}
\be
v = \sqrt{ QE / 2} \left[ { p_t \over Q E} + ( t + { k_0 \over Q E})\right]=
\sqrt{\frac m {2A}}e^{As}\qquad .
\label{tpt}
\ee
Since the parabolic cylinder functions have the property:
\be
{\cal{D}}_{\epsilon _M}^*[\lambda]=\alpha_M {\cal{D}}_{\epsilon _M}[-\lambda]
+\beta_M {\cal{D}}_{\epsilon _M}^*[-\lambda]\label{3DPar}
\ee
the in- and out- modes are related by the linear 
transformation:
\begin{eqnarray}
\label{Bogo3Psi}
\psi _{M,k}^{p\,out}&=&\alpha _M \/\psi _{M,k}^{p\,in}+\beta _M \/ (\psi
_{M,-k}^{a\,in})^* \nonumber \\
\psi _{M,k}^{p\,in}&=&\alpha _M^* \/\psi _{M,k}^{p\,out}-\beta _M \/ (\psi
_{M,-k}^{a\,out})^* \label{BogBog}
\end{eqnarray}
where
\begin{equation}
\alpha_M =  \frac{\sqrt{2\pi}e^{-i\frac{\pi}4}e^{-\frac {\pi}2
\epsilon_M}}{\Gamma(\frac12
+i\epsilon_M)} \qquad \mbox{\rm and } \qquad \beta_M= i e^{-\pi\epsilon_M}
\quad .
\label{coefBogo}
 \end{equation}
These  Bogoljubov coefficients are $k$ independent, but mass and charge
dependent.
The second quantized field $\Psi_M$ should be decomposed in both the in
or out basis
\begin{eqnarray}
\Psi _M(t,x)=\int dk\left[ \psi _{M,k}^{p\,{\left\{  in \atop  out 
\right.} }(t,x)
\ a_M^{{\left\{  in \atop  out  \right.} }(k)+(\psi _{M,k}^{a{\left\{  in
\atop  out  \right.}
}(t,x))^{*}\ b_M^{\dagger {\left\{  in \atop  out  \right.} }(k)\right] 
,\label{champ}
\end{eqnarray}
to define the in and out operators. From 
eq. (\ref{Bogo3Psi}) we obtain
\begin{eqnarray}
a^{in}_k&=&\alpha _M \ a^{out}_k-\beta _M \ b^{out\,\dagger }_{-k},\\ \nonumber
b^{in}_k&=&\alpha _M \ b^{out}_k-\beta _M \ a^{out\,\dagger }_{-k}.
\label{BogoOp}\end{eqnarray}
The Heisenberg state $\mid 0,in\rangle$ contains no particles at early
times, i.e.  it is annihilated by the in destruction operators.
At late times it contains pairs of particles, as expressed by the
relation
\begin{eqnarray}
\mid 0,in\rangle_M&=&{\cal{N}}^{-1}_M e^{\frac {\beta_M}{ \alpha_M }\int a^{out
\dagger}_kb^{out
\dagger}_{-k}\, dk}\mid 0,out\rangle_M \label{vacvac} 
\end{eqnarray}
where 
\begin{equation}
{\cal{N}}_M^{-1}=\/ _M\langle 0,out\mid 0,in\rangle_M. \label{overlaps}
\end{equation}
 is a normalization factor.
The mean number of created particles is 
\be 
N_{M} =\/ _M\bra{0, in} b^{out\,\dagger }_{-k} b^{out}_{-k}
\ket{0, in}_M = \vert \beta_M \vert^2 = e^{- 2 \pi M^2 / QE}\qquad.
\label{Sc}
\ee

\section{Particle interactions: emission of a neutral particle}
In this section we consider a uniformly accelerated detector
interacting with a neutral scalar field $\Phi_\mu$ of mass $\mu$.
We shall show that eq. (\ref{Rates2}), i.e. the relation between 
Schwinger and radiative processes which was obtained in \cite{ScUn} for
a massless field, still holds when the exchanged quanta are massive.
Moreover, we shall see that all amplitudes linear in the coupling constant
 can be expressed in terms of a single amplitude describing the creation 
of a pair of charged quanta.
The techniques developed in this section will be generalized in the next
section where  the field $\Phi_\mu$ is both
massive and charged.

The detector is described by  two charged scalar fields $\Psi_M$
and $\Psi_m$, with mass $M$
and $m$ and the same charge $Q$, propagating in the electric field
$E$. As in \cite{bps,rec,ScUn}, 
the interaction between the fields is supposed to be given by:
\begin{eqnarray}
H^{int}=g\int dx\left( \Psi _M^{\dagger }\Psi _m +
\Psi _M\Psi _m^{\dagger
}\right) \Phi _{\mu} \label{hamint} 
\end{eqnarray}
A first amplitude of interest is the  
 amplitude ${\cal{A}}$ of transition from an $in$ M-particle of momentum $k$
into an $out$ m-particle of momentum $k'$
and a $\mu$-particle of momentum $k''$, 
which we schematically write as
$\mbox{ M}(k)\ \to  \ \mbox{ m}(k')\ +\ \mu(k'')$.
It corresponds to spontaneous deexcitation of the detector (since $M>m$).
In the interaction representation, to first order in $g$, it is given by
\begin{eqnarray}
{\cal{A}}(k \vert k',k'') 
&=& -ig\int dtdx\langle
0,out\mid a_\mu (k^{\prime \prime })\ a_m^{out}(k^{\prime })\ \Psi
_M\Psi _m^{\dagger }\Phi
_\mu \ a_M^{in\dagger }(k)\mid 0,in\rangle  \nonumber\\
\label{akkk}
\end{eqnarray}
The factorization of the vacuum states as tensor product of three
vacua:
\begin{equation}
\mid 0,{\left\{ \strut^{in}_{out}  \right.}\rangle=
\mid 0,{\left\{ \strut^{in}_{out}   \right.}\rangle_M \otimes 
\mid 0,{\left\{ \strut^{in}_{out}  \right.}\rangle_m \otimes 
\mid 0 \rangle_\mu
\end{equation}
 leads to the
expression (see \cite{Niki2,Niki}): 
\begin{eqnarray} 
{\cal{A}}(k\vert k^{\prime },k^{{\prime }{\prime }})=
\frac 1{\alpha _M{\cal{N}}_M}\frac 1{\alpha _m{\cal{N}}_m}
(-i)g\int dt\,dx\ \psi _{Mk}^{p\,out}\psi _{mk^{\prime }}^{p\,in\,*}\phi
_{\mu k^{\prime \prime
}}^{*} \label{GenA}
\end{eqnarray}
where ${\cal{N}}_M$, ${\cal{N}}_m$ are the
overlaps
(see eqs. (\ref{vacvac},\ref{overlaps})).

At this point it is intersting to note that 
the quantity ${\cal{A}}(k \vert k',k'')$ describes
several different processes in addition to eq. (\ref{akkk}). Indeed it
is not difficult to verify using eqs. (\ref{modepin}--\ref{modeaout})
that the four transitions
\begin{eqnarray}
&
\mbox{M}(k)\ \to  \ \mbox{m}(k')\ +\ \mu (k'')&\quad ,\nonumber\\
&
\bar M(-k)\ \to  \ \bar m(-k')\ +\ {\mu} (-k'')&\quad , \nonumber\\
&
\bar m(k')\ +\ {\mu} (k'') \ \to \ \bar M(k)&\quad ,\nonumber\\
&
m(-k')\ +\ \mu(-k'')\ \to \ M(-k)\quad , \label{process4}
\end{eqnarray}
where $\bar M$ and $\bar m$ denote
antiparticles all have the same amplitude ${\cal{A}}(k \vert k',k'')$.
These different amplitudes are related by combinations of substituting
particles for antiparticles, changing the sign of the momentum $k \to
-k$
and permutting incoming and outgoing quanta. Their origin will
be further explained in the next section. Similar properties will
obtain for all the amplitudes we shall introduce. Thus the notation we
use for ${\cal{A}}(k \vert k',k'')$, and which is also used for all
the other amplitudes in this paper, is that the $|$ separates incoming
from outgoing quanta, and that $\pm k$ is the momentum of the $\Psi_M$
quanta,
$\pm k'$ of the $\Psi_m$ quanta, and $\pm k''$ of the $\Phi_\mu$ quanta.

Similarly, the amplitude ${\cal{B}}$ of spontaneous excitation 
of an $m$ particle into a $M$ particle
accompanied by the emission of a $\mu$ particle 
is given by
\footnotesize
\begin{eqnarray}
{\cal{B}}(k^{\prime }\vert k, -k^{{\prime }{\prime }})&=&-ig\int dtdx\langle
0,out\mid a_\mu(-k^{\prime \prime })\ a_M^{out}(k)\ \Psi _M^{\dagger}
\Psi _m 
\Phi _\mu 
\ a_m^{in\dagger}(k^{\prime})\mid 0,in\rangle\nonumber\\   
&=&\frac 1{\alpha _M {\cal{N}}_M}\frac 1{\alpha _m
{\cal{N}}_m}(-ig)\int dt\,dx\ 
\psi_{M,k}^{p\,in\,*}
\psi _{m,k^{\prime }}^{p\,out}
\phi _{\mu,-k^{\prime \prime }}^{*}\qquad . 
\end{eqnarray}
\normalsize

Due to the uniform acceleration, this spontaneous excitation amplitude
is non vanishing. As in the Unruh treatment, 
the ratio of the rates of spontaneous excitation to
spontaneous deexcitation is simply given by
\be
{ R_{m \to M} \over R_{M \to m}} = { \vert {\cal B} \vert ^2 \over
\vert {\cal A } \vert ^2 }.
\label{B/A}
\ee
since the norm of the amplitudes are independent of the 
momenta $k$ and $k'$.

Thus, when the detector reaches equilibrium, 
 the ratio of the probabilities to find the detector in its
excited or ground state are 
\be
{ P_{M\ equil} \over P_{m\ equil}}= { R_{m \to M} \over R_{M \to m}} = {
\vert {\cal B} \vert ^2 \over
\vert {\cal A } \vert ^2 }.
\ee
Our task is to calculate this ratio and to confirm the relation
with the Schwinger process, eq. (\ref{Sc}).
To this end we introduce a third amplitude ${\cal{V}}$
corresponding to the creation from vacuum  of  
an out-M-anti-particle, an out-m-particle and a $\mu$-particle. 
To first order in 
 $g$, it is given by
\begin{eqnarray}
{\cal{V}}(\vert -k,k^{\prime },k^{{\prime }{\prime }})&=&-ig\int dtdx\langle
0,out\mid a_\mu(k^{\prime \prime })\ a_m
^{out}(k^{\prime})\ b_M^{out}(-k)\ \Psi _M \Psi _m \Phi _\mu ^{\dagger} 
\mid 0,in\rangle  \nonumber \\
&=& \frac 1{\alpha _M {\cal{N}}_M}\frac 1{\alpha _m
{\cal{N}}_m}(-ig)\int dt\,dx\ 
\psi_{M,-k}^{a\,in\, *}
\psi _{m, k^{\prime }}^{p\,in\,*}
\phi _{\mu, k^{\prime \prime }}^{*}\qquad .
 \label{GenV} 
\end{eqnarray}
Using the Bogoljubov transformation for the $M$-particle wave function,
eq (\ref{BogBog}),
we can reexpress ${\cal V}$ in terms of the ${\cal A}$ amplitude
\begin{eqnarray}
{\cal{V}}(\vert -k,k^{\prime },k^{{\prime }{\prime }})&=&
\frac {-ig}{\alpha _M {\cal{N}}_M}\frac 1{\alpha _m
{\cal{N}}_m}\int dt\,dx\ 
\left[ \alpha_M \psi_{M,-k}^{a\,out\, *} + \beta_M \psi_{M,k}^{p\,out}
\right] \psi _{m k^{\prime }}^{p\,in\,*}
\phi _{\mu\, k^{\prime \prime }}^{*} \nonumber \\
&=& \beta_M {\cal{A}}(k\vert k^{\prime },k^{{\prime }{\prime }})
+  \alpha_M {\cal I}(k, k^{\prime },k^{{\prime }{\prime }})
\end{eqnarray}
Similarly,
using the Bogoljubov transformation for the $m$-particle 
we can reexpress ${\cal V}$ in terms of the ${\cal B}$ amplitude
\begin{eqnarray}
{\cal{V}}(\vert k, k^{\prime },k^{{\prime }{\prime }})&=&
 \beta_m {\cal{B}}(k'\vert k,-k^{{\prime }{\prime }})
+  \alpha_m {\cal I}^{\prime}(k, k^{\prime },k^{{\prime }{\prime }})\qquad.
\end{eqnarray}
Using the  identities
(which constitute the main mathematical result of this section, and are
proved hereafter)~:
\be
{\cal I} ={\cal I}^{\prime} =0 
\label{calI}\ee 
 one obtains
\be
\beta_m {\cal B}(k'|k,-k'') 
= {\cal V}(|-k,k',k'') = \beta_M {\cal A}(k|k',k'')\qquad ,
\label{BVA}
\ee
relations that we have depicted schematically in figure {\bf 1}.
Inserting eq. (\ref{BVA}) into eq.(\ref{B/A}) yields the link
between radiative and Schwinger processes
discussed in the introduction eq. (\ref{Rates2})
\be
{ P_{M\ equil} \over P_{m\ equil}}= { N_{M} \over N_{m}}
= \mbox{\rm{e}}^{-\pi(\frac{M^2-m^2}{QE})} .
\ee
\begin{figure}
\begin{center}
\resizebox{14cm}{!}{
   \includegraphics{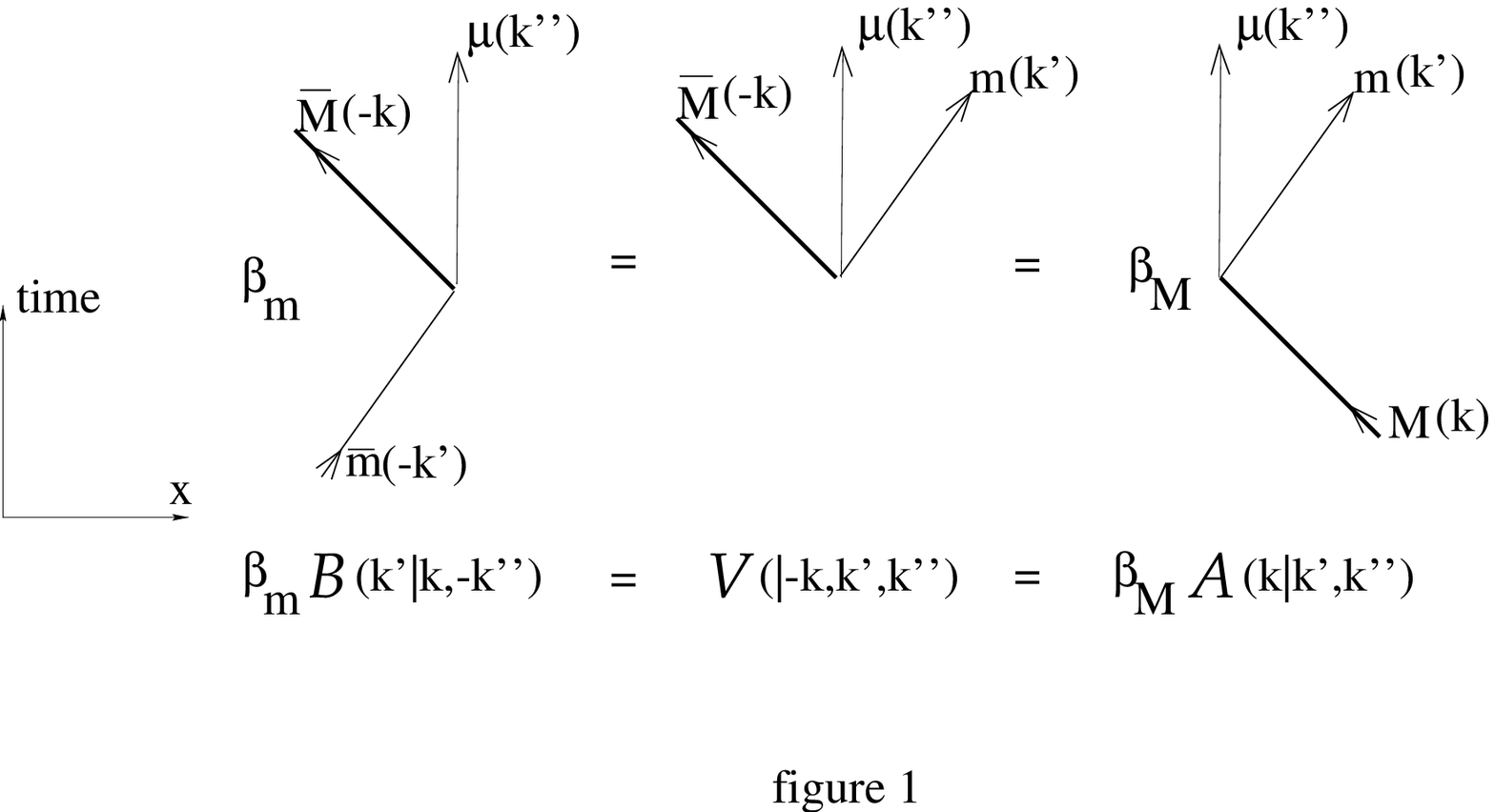}
   }
\end{center}
\caption{A graphic representation of the relation between amplitudes 
${\cal B}$, $ {\cal V}$, $ {\cal A}$ of section 3. 
Particles are denoted by $M(k)$, $m(k')$, $\mu(k'')$ 
according to their mass and momentum, and 
antiparticles by $\bar M$, $\bar m$ 
(their is no antiparticle for the $\Phi_\mu$ field which is real).
The momenta in each diagram are always conserved
 $k = k'+k''$.
The coefficients $\beta_M$, $\beta_m$ which weight the different
amplitudes are the amplitudes to produce a pair of $M,\bar M$ or $m,
\bar m$ particles from vacuum.
The picture is oriented both in
time and space: 
particles (antiparticles) are incoming from, and outgoing to, the
right (left)
according to 
their acceleration in the electric field. The chargless particles
$\mu$ are represented as vertical lines because they are not accelerated. }
\end{figure}

It remains to establish eq.(\ref{calI}).
To this end, we write
\ba
{\cal I}(k, k^{\prime },k^{{\prime }{\prime }})
&=& Cst\ \int dx dt
e^{ikx} e^{- i k^{\prime } x} e ^{ -i k^{{\prime }{\prime }} x}
 {\cal{D}}_{\epsilon _M}[-\lambda]
{\cal{D}}_{\epsilon _m}^*[\lambda^{\prime }]
e^{i\omega^{{\prime }{\prime }} t}
\nonumber\\
&=& 2\pi \delta(k  -k^{\prime }-k^{{\prime }{\prime }})
\int dt 
{\cal{D}}_{\epsilon _M}[-\lambda]
{\cal{D}}_{\epsilon _m}^*[\lambda^{\prime }]
e^{i\omega^{{\prime }{\prime }} t}
\ea
where we have introduced $\lambda = ( t + k/ QE)$,
$\lambda^\prime = ( t + k^\prime / QE)$.
To perform the $t$ integral we replace the Whittaker's functions
${\cal{D}}$ by their
integral representations eq. (\ref{DIrep}) to obtain
\ba
&{\cal I}(k, k^{\prime },k^{{\prime }{\prime }})&
= Cst^\prime\ \delta(k  -k^{\prime }-k^{{\prime }{\prime }})
\int dt
e^{\frac i2QE(t+\frac{k}QE)^2}
\ e^{-\frac
i2QE(t+\frac{k^{\prime }}QE)^2} e^{i\omega^{\prime\prime}t} \label{Ikk}\\
 \nonumber &&\int_0^\infty dudv\ 
e^{+iu\sqrt{2QE}(t+\frac{k}QE)+iu^2/2} u^{^{-i\epsilon _M-1/2}}
e^{+i(t+\frac{k^{\prime}}QE)\sqrt{2QE}v-iv^2/2}v^{i\epsilon _m-1/2} 
\end{eqnarray} 
The quadratic phases in $t$ cancel, and the $t$ integral yields
$\delta(u+v+
\frac{\omega^{{\prime}{\prime}}+k^{{\prime}{\prime}}}{\sqrt{2QE}}))$.
The  argument of the
delta function never vanishes on the domain of integration of $u$ and
$v$ since $\omega^{{\prime}{\prime}}+k^{{\prime}{\prime}} \neq 0$ for $\mu
\neq 0$. 
Hence ${\cal I} =0$. A similar reasoning shows that ${\cal I}^\prime=0$.


The interested reader will now find the calculation of the amplitudes 
themselves. By virtue of eq. (\ref{BVA}), we only need to calculate the
amplitude ${\cal V}$.
We start from eq.(\ref{GenV}) in which we reexpress
$(\Psi^{a\, in}_{M,-k})^*$ as $[(\Psi^{a\, out}_{M,-k})^* +\beta_M
 (\Psi^{p\, in}_{M,k})]/\alpha^*_M$. The first term does not contribute
 since it
 is equal to ${\cal I}(k,k',k'')/\vert\alpha\vert^2 $. The second term gives:
\footnotesize
\ba
{\cal V}(\vert -k,k^\prime, k^{\prime \prime})
&=& Cst\ \int dx dt
e^{ikx} e^{- i k^{\prime } x} e ^{ -i k^{{\prime }{\prime }} x}
 {\cal{D}}_{\epsilon _M}[\lambda]
{\cal{D}}_{\epsilon _m}^*[\lambda^{\prime }]
e^{i\omega^{{\prime }{\prime }} t}
\nonumber\\
&=& Cst^\prime\ \delta(k  -k^{\prime }-k^{{\prime }{\prime }})
\int dt\ 
e^{\frac
i2QE(t+\frac{k}QE)^2}
e^{-\frac i2QE(t+\frac{k^{\prime }}QE)^2}
e^{i\omega^{\prime\prime}t}
\nonumber\\
&&\int_0^\infty dudv
\ e^{-iu\sqrt{2QE}(t+\frac{k}QE)+iu^2/2}u^{^{-i\epsilon _M-1/2}}
e^{i(t+\frac{k^{\prime
}}QE)\sqrt{2QE}v-iv^2/2}v^{i\epsilon _m-1/2} 
\nonumber\\
&=& Cst^{\prime\prime}\ \delta(k  -k^{\prime }-k^{{\prime }{\prime }})
\int_0^\infty dudv\ \delta(-u+v+
\frac{\omega^{{\prime}{\prime}}+k^{{\prime}{\prime}}}{\sqrt{2QE}}))
\\ \nonumber 
&&
e^{i(\frac{k^{\prime
}}QE)\sqrt{2QE}v-iv^2/2}v^{i\epsilon _m-1/2}u^{^{-i\epsilon _M-1/2}}
e^{-iu\sqrt{2QE}(\frac{k}QE)+iu^2/2} 
\nonumber\\
&=& Cst^{\prime\prime\prime}\ \delta(k  -k^{\prime }-k^{{\prime }{\prime }})
\int_0^\infty dv\ e^{iv\frac{\omega
-k^{\prime \prime}}{\sqrt{2QE}}}v^{i\epsilon _m-1/2}(v+\frac{\omega
+k^{\prime \prime}}
{\sqrt{2E}})^{{-i\epsilon _M-1/2}}\nonumber
\end{eqnarray}
\normalsize 
The last integral gives an integral representation of a Whittaker's
function (See \cite{GR}, formula 3.383.4). Reinstating the value of
$Cst^{\prime\prime\prime}$ yields 
\begin{eqnarray} 
&{\cal{V}}(\vert -k,k^{\prime },k^{{\prime }{\prime }})=g\frac 1{\alpha
_M{\cal{N}}_M \alpha _m{\cal{N}}_m}
\delta (k-k^{\prime }-k^{\prime \prime})
\frac{e^{-\frac{3\pi} 4 \epsilon _m}e^{-\frac {3 \pi}
4\epsilon_M}}{2QE \/\sqrt{2\omega }}&
\\
& \frac1{\mu} 
e^{-\frac i{2QE}(k +k^{\prime })\omega} 
 \left( \frac{\omega-k^{\prime \prime}}{\omega+k^{\prime \prime}} 
\right)^{\frac i2(\epsilon _M-\epsilon _m)}
W_{-\frac i2(\epsilon _m+\epsilon _M),\frac i2(\epsilon
_m-\epsilon _M)}\left[{-\frac{i\mu ^2}{2QE}}\right] &\nonumber
\label{Aneutral}
\end{eqnarray}  
It is interesting to note that the delta function $
\delta(-u+v+
\frac{\omega^{{\prime}{\prime}}+k^{{\prime}{\prime}}}{\sqrt{2QE}})$
has the interpretation of ensuring local energy conservation. Indeed,
replacing $u$ and $v$ by their classical relation to $t$ and $p_t$ 
eq. (\ref{tpt}),
yields
simply $-p_t + p_t^\prime + \omega =0$ where $p_t$
is the energy of the M-particle, $ p_t^\prime$ of the m-particle, and
$\omega$ of the $\mu$ particle.

We conclude this section by calculating the $\mu \to 0$ limit of the above
amplitude. This will allow us to make contact with the expressions
of \cite{ScUn} that were obtained using a different method. 
The evaluation of this limit
needs some precautions. Indeed for $k''<0$, the support of the delta
function given by the
 integral over $t$ in eq.(\ref{Ikk}) rejoins the boundary of the domain of
integration. In
order to avoid  ambiguities, we have to substitute in eq.(\ref{GenV}),
according to the sign of
$k''$,  the decomposition of $(\Psi^{a\, in}_{M,-k})^*$  or $(\Psi^{p\,
in}_{m,k'})^*$ in terms
of $in$ and $out$ fields and take into account the infinitesimal imaginary
part that the squared
masses of the fields share.
For small value of $\mu$ one finds:
\footnotesize
\begin{eqnarray}
&{\cal{V}}(\vert -k,k^{\prime },k^{{\prime }{\prime }})
&=_{\mu \to 0}g\frac 1{\alpha
_M{\cal{N}}_M \alpha _m{\cal{N}}_m}
\delta (k-k^{\prime }-k^{\prime \prime})
\frac {e^{-\frac{\pi}{2} (\epsilon _m-\epsilon_M)}}{2QE \sqrt{2\omega }}
\frac{e^{3i\pi/4}}{\sqrt{2QE}} e^{-\frac i{2QE}(k +k^{\prime })\omega}
 \nonumber\\
&&\left\{ \theta(k^{\prime \prime})
(\frac {2k''}{QE})^{\frac i2
  (\epsilon_m-\epsilon_M)}
\frac 1{\Gamma(\frac 12+i\epsilon_M)\Gamma(1+i\epsilon_m-i\epsilon_M)}\right.
\nonumber \\
&&\ -\left.
\theta(-k^{\prime \prime})e^{\frac \pi 2 (\epsilon_m-\epsilon_M)}
(\frac {2{k^{\prime \prime}}}{QE})^{\frac i2
  (\epsilon_M-\epsilon_m)}
\frac {1}{\Gamma
(\frac12+i\epsilon_m)\Gamma(1-i\epsilon_m+i\epsilon_M)} \right \}
\end{eqnarray}
\normalsize
 Note the surprising fact that in this limit, the amplitudes of decay of
the same  process  
but with the opposite momenta differ. However, this is peculiar to two
dimensions. Indeed
in higher dimension, the mass $\mu$  contains the squared transversal
momentum and so 
even in the massless limit, vanishes
only on a domain of zero measure in phase space. 
\section{Particle interactions: emission of a charged particle}
In this section we shall suppose that the $\Phi_\mu$ field is also
charged. Thus there are three fields $\Psi_M$, $\Psi_m$, $\Phi_\mu$,
with masses $M$, $m$, $\mu$ and charges $Q$, $q$, $\alpha$ respectively.
These three fields interact through the hamiltonian
\begin{eqnarray}
H^{int}=g\int dx\left( \Psi _M^{\dagger }\Psi _m\Phi
_{\mu}+\Psi _M\Psi _m^{\dagger
}\Phi _{\mu}^{\dagger }\right) \label{hamint2}\qquad .
\end{eqnarray}
Charge conservation requires
\be
Q= q+\alpha \qquad.
\label{chargecons}
\ee
The amplitude ${\cal{A}}$ of transition from an $in$ M-particle
into an $out$ m-particle
and a $out$ $\mu$-particle
is given at first order in perturbation theory by: 
\begin{eqnarray} 
{\cal{A}}(k|k',k'')&=&\langle 0,out\mid
a_\mu ^{out}(k^{\prime \prime })\ a_m^{out}(k^{\prime })
(-i)\int dt H^{int} a_M^{in\dagger
}(k)\mid 0,in\rangle \nonumber\\
&=&
\frac 1{\alpha _M{\cal{N}}_M}\frac 1{\alpha _m{\cal{N}}_m}\frac 1{\alpha
_\mu {\cal{N}}_\mu
}(-i)g\int dt\,dx\ \psi _{Mk}^{p\,out}\psi _{mk^{\prime }}^{p\,in\,*}\phi
_{\mu k^{\prime \prime
}}^{p\,in\,*}\nonumber\\
& =& {\mbox{\rm C}}\,
\delta (k-k^{\prime }-k^{{\prime }{\prime }}) \Ar
\end{eqnarray}
where 
\be
{\mbox{\rm C}} = {-ig \over \sqrt{ 2 \pi}}
\frac 1{{\alpha _M{\cal{N}}_M}
{\alpha _m{\cal{N}}_m}
{\alpha _\mu {\cal{N}}_\mu}}
\ee
and we have introduced the reduced amplitude 
\begin{equation}
\Ar=\int_{-\infty}^{+\infty} dt {\cal{D}}_{\epsilon _M}^{*}[-\lambda]\,
{\cal{D}}_{\epsilon _m}^{*}[\lambda^{\prime}]{\cal{D}}_{\epsilon
_{\mu}}^{*}[\lambda^{\prime \prime}] \label{A3D}
\end{equation} 
and used the abbreviated notation for the parabolic cylinder functions
introduced in eq. (\ref{Par}). Note also the appearance of the factor 
$1/{\alpha _\mu {\cal{N}}_\mu}$ which arises due to the vacuum
instability of the $\mu$ field.

As in the previous section, the processes
\begin{eqnarray}
&
\mbox{M}(k)\ \to  \ \mbox{m}(k')\ +\ \mu (k'')&\quad ,\nonumber\\
&
\bar M(-k)\ \to  \ \bar m(-k')\ +\ \bar {\mu} (-k'')&\quad , \nonumber\\
&
\bar m(k')\ +\ \bar {\mu} (k'') \ \to \ \bar M(k)\quad , \nonumber\\
&m(-k')\ +\ \mu(-k'')\ \to \ M(-k)&\quad ,
\label{process4B}
\end{eqnarray}
all have the same amplitude ${\cal{A}}(k|k',k'')$. Note that these
differ slightly from eq. (\ref{process4})  because one has to distinguish
between $\mu $ and $\bar \mu$ since the field $\Phi_\mu$ is charged.

Similarly the amplitude ${\cal B}$ for a anti-m-particle to
spontaneously excite into a anti-M-particle and a $\mu$-particle is
\begin{eqnarray}
{\cal{B}}(k^{\prime }\vert k,-k^{{\prime }{\prime }})&=&-i\int dtdx\langle
0,out\mid a_\mu
^{out}(k^{\prime \prime })\ b_M^{out}(-k)\ \Psi _M \Psi _m^\dagger 
\Phi _\mu^\dagger 
\ b_m^{in\dagger
}(-k^{\prime})\mid 0,in\rangle  \nonumber \\
&=&{\mbox{\rm C}}\,
\delta (k-k^{\prime }-k^{{\prime }{\prime }}) \Br
\label{GenB22} 
\end{eqnarray}
where
\ba
\Br&=&\int_{-\infty}^{+\infty}dt\/
{\cal{D}}_{\epsilon_M}^{*}[\lambda]{\cal{D}}_{\epsilon_m}^{*}[-\lambda^{\prime}]
{\cal{D}}_{\epsilon_{\mu}}^{*}[\lambda^{\prime \prime}] \label{B3D}
\ea
Another amplitude that we need to introduce is the 
amplitude ${\cal C}$ for a anti-$\mu$-particle to
spontaneously excite into a anti-M-particle and a m-particle:
\begin{eqnarray}
{\cal{C}}(k^{{\prime }{\prime }} \vert -k^{\prime },k)&=&-i\int dtdx\langle
0,out\mid b_m
^{out}(-k^{ \prime })\ a_M^{out}(k)\ \Psi _M^{\dagger} \Psi _m \Phi _\mu 
\ a_\mu^{in\dagger
}(k^{\prime\prime})\mid 0,in\rangle  \nonumber \\
&=&{\mbox{\rm C}}\,
\delta (k-k^{\prime }-k^{{\prime }{\prime }}) \Cr
\label{GenC2} 
\end{eqnarray}
where
\ba
\Cr&=&\int_{-\infty}^{+\infty}dt\/
{\cal{D}}_{\epsilon_M}^{*}[\lambda]{\cal{D}}_{\epsilon_m}^{*}[\lambda^{\prime}]
{\cal{D}}_{\epsilon_{\mu}}^{*}[-\lambda^{\prime \prime}] \qquad ,
\ea
and the amplitude ${\cal V}$ for spontaneous creation from vacuum of
an anti-M-particle, a m-particle and a $\mu$-particle
\begin{eqnarray}
{\cal{V}}(\vert -k,k^{\prime },k^{{\prime }{\prime }})&=&-i\int dtdx\langle
0,out\mid a_m
^{out}(k^{ \prime })\ b_M^{out}(-k)\ a_\mu^{out}
(k^{\prime\prime})\Psi _M^{\dagger} \Psi _m \Phi _\mu 
\ \mid 0,in\rangle  \nonumber \\
&=&{\mbox{\rm C}}\,
\delta (k-k^{\prime }-k^{{\prime }{\prime }}) \Vr
\label{GenV2} 
\end{eqnarray}
where
\ba
\Vr&=&\int_{-\infty}^{+\infty}dt\/
{\cal{D}}_{\epsilon_M}^{*}[\lambda]{\cal{D}}_{\epsilon_m}^{*}[\lambda^{\prime}]
{\cal{D}}_{\epsilon_{\mu}}^{*}[\lambda^{\prime \prime}] \label{V3D}
\ea

The rules governing the product of ${\cal D}$ functions in the reduced
amplitudes {\bf A, B, C, V} can be summarized as:
\begin{itemize}
\item The sum of the momenta is conserved. We always assume that  
$k=k^{\prime}+k^{\prime \prime}$.
\item Equations (\ref{modepin} and \ref{modeain}) tell us that a function
${\cal{D}}_{\epsilon _M}^{*}[-\lambda]$ is associated to an incoming
particle (of mass $M$, charge $Q>0$, and momentum $k$)
or an incoming antiparticle (of
mass $M$, charge $-Q$, and momentum $-k$) 
\item 
Equations (\ref{modepout} and \ref{modeaout}) tell us that a function
${\cal{D}}_{\epsilon _M}^{*}[\lambda]$ is associated to an outgoing
particle (of mass $M$, charge $Q>0$, and momentum $k$)
or an outgoing antiparticle (of
mass $M$, charge $-Q$, and momentum $-k$)
\item The change of variable $t \to -t$ in the integrals accompagnied
  with a change in sign of momentum $k \to -k$ replaces 
incoming quanta by outgoing quanta and {\it vice versa}, without
  changing the value of the integral.
\end{itemize}
These rules give directly the relation eq. (\ref{process4B}) and similar ones
for ${\bf B}$, ${\bf C}$ and ${\bf V}$. Therefore the 64 possible
amplitudes describing first order interactions between $\Psi_M$,
$\Psi_m$ and $\Phi_\mu$ quanta can all equal to either 
{\bf A, B, C, V}\footnote{Note that 
these amplitudes are all integrals of products of three
${\cal D}^*$ functions. Mathematically one could also consider
intergals of products of three ${\cal D}^*$ and ${\cal D}$ functions. The 
Bogoljubov transformation (\ref{Bogo3Psi}) ensures that they can be 
reduced to a combinaison of
{\bf A, B, C, V}}.
\begin{figure}
\begin{center}
\resizebox{14cm}{!}{
   \includegraphics{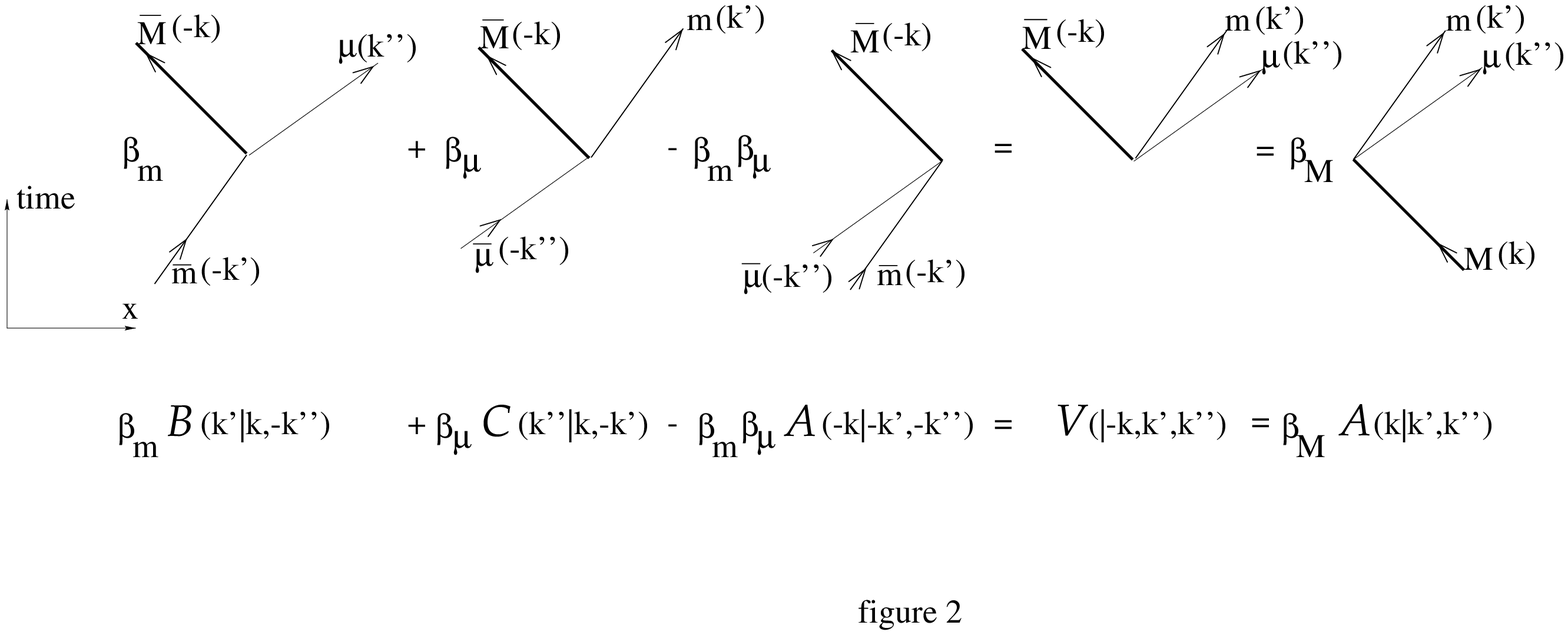}
   }
\end{center}
\caption{
A graphic representation of the relation between the amplitudes 
${\bf A}$, ${\bf B}$, ${\bf C}$, ${\bf V}$ when the field $\Phi_\mu$
is charged. The relation between the amplitudes is more complicated
than in figure 1 because of the non vanishing amplitude $\beta_\mu$ to
produce
pairs of $\mu, \bar \mu$ quanta.
The conventions are the same as in figure 1, except that
$\mu$ and $\bar \mu$ denote a particle and antiparticle 
of the $\Phi_\mu$ field, and are
represented by
slanted lines since they are also
accelerated by the electric field. 
Note that 
this picture is not the unique representation of the amplitudes 
${\bf A}$, ${\bf B}$, ${\bf C}$, ${\bf V}$ since as indicated in the
text each amplitude corresponds to 4 different processes. We have used this
to represent the amplitude ${\bf A}(-k\vert -k',-k'')$ as the amplitude 
$\bar m(k') + \bar \mu(k'') \to \bar M(k)$, but it could equally well be
represented by $M(-k) \to m(-k') +\mu(-k'')$. 
}
\end{figure}

There is a further identity which will play a crucial r\^ole, namely
 \begin {equation} {\cal I}=
\int_{-\infty}^{+\infty}dt\/
{\cal{D}}_{\epsilon_M}[-\lambda]{\cal{D}}_{\epsilon_m}^{*}[\lambda^{\prime}]
{\cal{D}}_{\epsilon_{\mu}}^{*}[\lambda^{\prime \prime}]=0  \qquad .\label{key}
\end{equation}
It generalizes eq. (\ref{calI}) and it is proven also by using the integral
representation 
of the ${\cal D}$ functions given in eq. (\ref{DIrep}).
As in eq. (\ref{Ikk}), 
one verifies that ${\cal I}$ still
vanishes simply because due to charge conservation the quadratic phases in $t$
cancel each other and the remaining $t$ integral yields to a delta
function whose argument
is strictly positive on the domain of integration. 
This identity implies relations between the
amplitudes ${\cal A , B, C, V}$ which generalize eq. (\ref{BVA}).
Indeed from eqs (\ref{3DPar}) and (\ref{key}) we
obtain immediately: 
\begin{equation}
\Vr=\beta_M\/\Ar.
\label{VA}
\end{equation}
On the other hand, starting from eq. (\ref{V3D}) and using the identity
(\ref{3DPar}) to split into  sums, successively,  the functions
${\cal{D}}_{\epsilon_m}^{*}[\lambda^{\prime}],
{\cal{D}}_{\epsilon_{\mu}}^{*}[\lambda^{\prime \prime}]$ and
$\alpha_m {\cal{D}}_{\epsilon_m}^{*}[-\lambda^{\prime}]$  and finally using
the complex conjugate
version of relation (\ref{key}), we obtain:
\footnotesize
\begin{eqnarray}
\Vr&=&\beta_m\/\Br +\beta_{\mu}\Cr - \beta_{\mu}\beta_m \Arm
\label{VBCA}
\end{eqnarray}
\normalsize
These relations are illustrated in figure 2.
Using eq. (\ref{VA}) to eliminate $\Vr$ yields
\begin{eqnarray}
\frac \Br \Ar=\frac {\beta_M} {\beta_m} - \frac {\beta_\mu} {\beta_m} \frac
{\Cr} {\Ar} +\beta_\mu \frac {\Arm} {\Ar}
\label{VBC}
\end{eqnarray}
This shows that when the charge of the exchanged 
particle tends to zero, i.e.  $\beta_\mu=e^{- \pi \mu^2 / 2 \alpha} \to 0$, the
ratio of the amplitudes $\Ar$ and $\Br$ is still directly
related to the amplitudes for the Schwinger process:
\be
\frac \Br \Ar = \frac {\beta_M} {\beta_m}
+ O(\beta_\mu)
\label{relatedS}
\ee
This estimate is  proven in Appendix 2 
where the $\alpha \to 0$ limit is studied in detail. 
It is shown that while  each amplitude involving a charged
$\mu$--particle
involves polynomial corrections in
$\alpha/\mu$, the ratio of the population of $m$ and $M$ species is in
thermodynamical equilibrium up only to non--perturbative corrections,
as expressed in eq. (\ref{relatedS}).

Thus 
the equilibrium distribution of the detector states is given by
\ba
{P_{M\ equil} \over P_{m \ equil}} &=&
{|\beta_M|^2 \over |\beta_m|^2 }+ O(\beta_\mu)
\nonumber\\
&=& e^{- \pi { M^2 / Q E} + \pi { m^2 / q E}} + O(\beta_\mu)
\ea
Therefore this equilibrium distribution is governed by the
finite change of the horizon area, c.f. the Introduction.

Upon taking the double limit $\delta M/M \to 0, \alpha / Q \to 0$ with the
mean acceleration \be
\bar a = { ({Q + q\over 2})E  \over ({M + m \over 2})}
\ee
fixed, one gets the linearized expression
\be 
 {P_{M\ equil} \over P_{m \ equil}} 
= exp\left[- {2 \pi \over \bar a} (\Delta m -
\alpha {E \over 2 \bar a})
+ O(\alpha^2)\right]+ O(\beta_\mu)
\ee
governed by a temperature $\bar a /2 \pi$ and an electric potential
$(= E/ 2 \bar a)$. 

Thus when the charge of the emitted particle is small enough with respect to
its mass square (so as to neglect its vacuum instability), its effect, at the
linearized level, is
to modify Unruh equilibrium by the addition of an a electric potential
thereby enlarging the relation to thermodynamics in a non trivial way
\\

{\bf Acknowledgments:}
Cl. Gabriel, S. Massar and Ph. Spindel gratefully acknowledge  the Fonds
National de la Recherche
Scientifique for generous financial supports. S. Massar would also like to
acknowledge partial
support  from  grant 614/95 of the
Israel Science Foundation and the
Universit\'e de Mons-Hainaut, where part of this work was carried out,
for hospitality.


\section {Appendix 1: Amplitudes of decay due to the exchange of charged
particle}

For completeness we give here closed forms for the 
amplitudes {\bf A, B, C, V}.
We first discuss the
${\cal{A}}(k|k^{\prime },k^{{\prime }{\prime }})$ amplitude. We have
to 
evaluate the integral \ref{A3D}, which in terms of Whittaker's functions,
reads as~:  
\begin{eqnarray}
{\cal{I}}_1&=&\int dt \ D_{-i\epsilon _M-1/2} 
\left[-e^{-3i\pi /4}(t+\frac k {MA})\sqrt{2MA}\right]  \label{I3D} \\
&&D_{-i\epsilon _m-1/2}\left[ e^{-3i\pi /4}(t+\frac{k^{\prime
}}qE)\sqrt{2qE}\right]
D_{-i\epsilon _\mu -1/2}\left[ e^{-3i\pi /4}(t+\frac{k^{\prime \prime
}}{\mu \alpha })
\sqrt{2\mu \alpha }\right] \nonumber
\end{eqnarray}
The evaluation of this integral is similar to the pattern followed to
obtain eq.(\ref{GenV}). 
First we split the function $D_{-i\epsilon _M-1/2}\left[ e^{i\pi /4}(t+\frac
k{QE})\sqrt{2QE}\right] $ into two others
functions thanks to the relation (\ref{3DPar}) and reexpress 
all the parabolic cylinder functions involved in terms of their integral
representation (\ref{DIrep}). This yields: 
\begin{eqnarray} 
{\cal{I}}_1&=&
\frac {e^{-\frac{\pi \epsilon _m}4}e^{-\frac{\pi \epsilon _\mu }4}}
{\Gamma (i\epsilon _m+1/2)\Gamma (i\epsilon _\mu +1/2)}
\frac{e^{\frac{i\pi }4}}{\sqrt{2\pi }}\nonumber \\
&&\int dt\ e^{-\frac i2qE(t+\frac{k^{\prime }}{qE})^2}
e^{-\frac i2\alpha E (t+\frac{k^{\prime\prime}}{\alpha E })^2}
e^{\frac i2QE(t+\frac{k}{QE})^2} \nonumber \\
&&\int_0^\infty dudvdw\ e^{iv\sqrt{2qE}(t+\frac{k^{\prime }}{qE})-iv^2/2}
e^{iw\sqrt{2\alpha E }(t+\frac{k^{\prime\prime}}{\alpha E
})-iw^2/2}v^{i\epsilon
_m-1/2}w^{i\epsilon _\mu -1/2}u^{-i\epsilon _M-1/2} \nonumber \\ 
&&\left[ e^{\frac \pi 4 \epsilon
_M}e^{-\frac{3i\pi }8}e^{-iu\sqrt{2QE}(t+\frac{k}{QE})+iu^2/2}+
e^{-\frac{3\pi} 4 \epsilon
_M}e^{\frac{i\pi }8}e^{iu\sqrt{2QE}(t+\frac{k}{QE})+iu^2/2}\right]
\label{twotermsI1c} 
\end{eqnarray}
with $k=k^{\prime}+k^{\prime\prime}$.
Charge conservation $Q = q + \alpha$ eliminates the
quadratic terms in
$t$ and the $t$ integration leads to two delta functions~:
$\delta(\sqrt{qE}v+\sqrt{\alpha E}w\mp \sqrt{QE}u)$. The positivity of the
$u,v$ and $w$
variables makes only the first one contributing to the amplitude. 
After some elementary algebra, we obtain for ${\cal{I}}_1$ the expression:
\begin{eqnarray}
{\cal{I}}_1=\frac{e^{\frac \pi 4(\epsilon _M-\epsilon _m-\epsilon _\mu
)}e^{-\frac{i\pi }8}}{\Gamma (i\epsilon _m+1/2)\Gamma (i\epsilon _\mu
+1/2)}\frac{\sqrt{2\pi }}{\sqrt{2QE}}e^{\frac
i2(\frac{k^2}{QE}\,-\frac{k^{\prime \,2}}{qE}-
\frac{k^{\prime\prime}\,^2}{\alpha E })} {\cal{I}}_2.
\end{eqnarray} 
where:
\begin{eqnarray}
{\cal{I}}_2&=&\int_0^\infty dvdw\ e^{-\frac i{2QE}(\sqrt{\alpha E
}v-\sqrt{qE}w)^2+i\sqrt{2}\frac{QE \,k^{\prime }-qE\,k}{\sqrt{QEqE\alpha E
}}(\frac{\sqrt{\alpha E
}}{\sqrt{QE}}v-\frac{\sqrt{qE}}{\sqrt{QE}}w)}\\ \nonumber
&& \quad (\frac{\sqrt{\alpha E
}}{\sqrt{QE}}w+\frac{\sqrt{qE}}{\sqrt{QE}}v)^{-i\epsilon
_M-1/2}v^{i\epsilon _m-1/2}w^{i\epsilon _\mu -1/2}\\ \nonumber
&=&(\frac{\sqrt{QE}}{\sqrt{\alpha E }})^{i\epsilon
_m+1/2}
(\frac{\sqrt{QE}}{\sqrt{qE}})^{i\epsilon _\mu +1/2}
(\frac{\sqrt{\alpha E}}{\sqrt{qE}})^{i\epsilon _M +1/2}
\Gamma (i{\cal{E}}+\frac 12)e^{\frac{i\Omega ^2}2}e^{\frac{\pi
{\cal{E}}}4}e^{-\frac{i\pi
}8} {\cal{I}}_3 ,
\end{eqnarray} 
with 
\be\Omega=\frac{qE \,k-QE\, k^{\prime}} {\sqrt{QEqE\alpha E}}\ee 
and 
\be{\cal{E}}=\epsilon_m+\epsilon_\mu-\epsilon_M
\ee
and
\begin{eqnarray} 
{\cal{I}}_3=&&\left\{ 
D_{-i{\cal{E}}-\frac 12}\left[+ \sqrt{2}e^{i\frac{\pi} 4}\Omega \right]
B(i\epsilon _\mu +\frac
12,-i{\cal{E}}+\frac 12)\ \right. 
 \nonumber \\
&& \left. _2F_1(i\epsilon _M+\frac 12,i\epsilon _\mu +\frac 12,1+i\epsilon
_M-i\epsilon
_m;-\frac{\alpha E }{qE})   \right.  \nonumber \\
&& + \left. D_{-i{\cal{E}}-\frac 12}\left[
\sqrt{2}e^{-i\frac{\pi} 4}\Omega \right] 
B(i\epsilon _m+\frac 12,-i{\cal{E}}+\frac 12)\ \right.  \nonumber \\
 && \left. _2F_1(i\epsilon
_M+\frac 12,i\epsilon _m+\frac 12,1+i\epsilon _M-i\epsilon _\mu
;-\frac{qE}{\alpha E })\right\}.
\label{I3} 
\end{eqnarray} 
The way to obtain this last result and others similar is postponed to the
end of this appendix.
Collecting all these results, the amplitude $\Ar$ reads:
\begin{eqnarray}
&&\Ar  =\frac {\sqrt{2\pi}
e^{-\frac{i\pi }4}e^{-\frac \pi 2
(\epsilon_m+\epsilon_\mu)}
e^{\frac{\pi{\cal{E}}}8}
e^{\frac
i2(\frac{k^2}{QE}\,-\frac{k^{\prime\,2}}{qE}-\frac{k^{\prime\prime}\,^2}{\alpha
E })} e^{\frac{i\Omega ^2}2}} {
(2QE)^{1/4}(2qE)^{3/4}(2
\alpha E)^{1/4}
\Gamma (i\epsilon _m+1/2)\Gamma
(i\epsilon _\mu+1/2)}\nonumber \\
&&(\frac Qq)^{\frac i2\epsilon_{\mu}}
(\frac {\alpha}q)^{\frac i2\epsilon_M}
(\frac Q \alpha)^{\frac i2\epsilon_m}
\Gamma(i{\cal{E}}+\frac 12)
 {\cal{I}}_3
\label{Acharged} 
\end{eqnarray} 
As expected this result is symmetric with respect to the exchange
$(m,a,k^{\prime})\leftrightarrow
(\mu,\alpha,k^{\prime \prime})$.\\
The computation of the $\Br$
is somewhat different. 
Starting from the general expression (\ref{B3D}), we have to evaluate the
integral:
\begin{eqnarray}
{\cal{J}}_1&=&\int dt \ D_{-i\epsilon _M-1/2} 
\left[ e^{-3i\pi /4}(t+\frac k {QE})\sqrt{2QE}\right] \\ \nonumber
&&D_{-i\epsilon _m-1/2}\left[- e^{-3i\pi /4}(t+\frac{k^{\prime
}}{qE})\sqrt{2qE}\right]
D_{-i\epsilon _\mu -1/2}\left[ e^{-3i\pi /4}(t+\frac{k^{\prime \prime
}}{\alpha E })\sqrt{2\alpha E }\right] \nonumber
\end{eqnarray}
with $k=k^{\prime}+k^{\prime \prime}$.
\par\noindent
After the same transformations as those performed to evaluate 
${\cal{I}}_1$, we obtain:\\
\begin{eqnarray}
&&{\cal{J}}_1=
\frac {e^{-\frac \pi 4 (\epsilon _m+\epsilon _\mu)}}{\Gamma (i\epsilon
_m+1/2)}\frac
1{\Gamma (i\epsilon _\mu +1/2)} \frac{e^{\frac{i\pi }4}}{\sqrt{2\pi
}}\int dt\ e^{-\frac i2qE(t+\frac{k^{\prime }}{qE})^2} e^{-\frac i2\alpha E
(t+\frac{k^{\prime \prime}}{\alpha E })^2} e^{\frac
i2QE(t+\frac{k}{QE})^2} \nonumber \\
&&\int_0^\infty dudvdw\ e^{-iv\sqrt{2qE}(t+\frac{k^{\prime
}}{qE})-iv^2/2}e^{iw\sqrt{2\alpha E
}(t+\frac{k^{\prime\prime}}{\alpha E })-iw^2/2}v^{i\epsilon
_m-1/2}w^{i\epsilon _\mu
-1/2}u^{-i\epsilon _M-1/2} \nonumber \\
 &&\left[ e^{\frac \pi 4 \epsilon _M}e^{-\frac{3i\pi
}8}e^{iu\sqrt{2QE}(t+\frac{k}{QE})+iu^2/2}+e^{-\frac{3\pi} 4 \epsilon
_M}e^{\frac{i\pi
}8}e^{-iu\sqrt{2QE}(t+\frac{k}{QE})+iu^2/2}\right] . 
\end{eqnarray} 
As previously charge conservation implies that 
the quadratic terms in $t$ cancel and the
$t$-integration yields a delta
functions $\delta(\sqrt{\alpha E}w-\sqrt{qE}v\pm \sqrt{QE}u)$. However,
this time, both 
terms will contribute to ${\cal{J}}_1$ and we are left with the following
expression: \\
 \begin{eqnarray} 
{\cal{J}}_1&=&\frac{e^{-\frac {\pi} 4(\epsilon _m+\epsilon _\mu
)}\, e^{\frac{i\pi }4}}{\Gamma (i\epsilon _m+1/2)\Gamma (i\epsilon _\mu
+1/2)}\sqrt{2\pi
} e^{\frac i2\left\{ \frac{k^2}{QE}\,-\frac{k^{\prime
\,2}}{qE}-\frac{(k^{\prime\prime})^2}{\alpha E }\right\} } \nonumber \\
&& \left[ \frac 1 {\sqrt{2qE}} e^{\frac \pi 4
\epsilon_M}e^{-\frac{3i\pi}8} {\cal{J}}_2+\frac 1 {\sqrt{2\alpha
E}}e^{-\frac{3 \pi} 4
\epsilon_M}e^{\frac{i\pi}8}
{\cal{J}}_3 \right] \quad . 
\end{eqnarray} 
where 
\begin{eqnarray}
{\cal{J}}_2&=&\int_0^\infty dudw e^{-\frac i{2qE}(\sqrt{\alpha E
}u+\sqrt{QE}w)^2+i\sqrt{2}\frac{qE\,k-QE
\,k^{\prime}}{\sqrt{QEqE\alpha E }}(\frac{\sqrt{\alpha E }}
{\sqrt{q E}}u+\frac{\sqrt{QE}}{\sqrt{q E}}w)} \nonumber \\
&&(\frac{\sqrt{\alpha E
}}{\sqrt{q E}}w+\frac{\sqrt{QE}}{\sqrt{qE}}u)^{i\epsilon
_m-1/2}u^{-i\epsilon _M-1/2}w^{i\epsilon _\mu -1/2} \label{J2} 
\end{eqnarray}
and:
\begin{eqnarray}
{\cal{J}}_3&=&\int_0^\infty dudv e^{-\frac i{2\alpha E}(\sqrt{Q E
}v+\sqrt{qE}u)^2+i\sqrt{2}\frac{qE\,k-QE
\,k^{\prime}}{\sqrt{QEqE\alpha E }}(\frac{\sqrt{QE }}
{\sqrt{\alpha E}}v+\frac{\sqrt{qE}}{\sqrt{\alpha E}}u)}
\nonumber \\ 
&&(\frac{\sqrt{QE}}{\sqrt{\alpha E}}u+\frac{\sqrt{q E
}}{\sqrt{\alpha E}}v)^{i\epsilon
_\mu-1/2}v^{i\epsilon _m-1/2}u^{-i\epsilon _M -1/2}\label{J3} 
\end{eqnarray}
Note that the last integral ${\cal{J}}_3$ can be obtained from the first 
one ${\cal{J}}_2$ by
exchanging
$(m,\ a)$ with $(\mu,\ \alpha)$ and $(k,\ k^{\prime},\ 
k^{\prime \prime})$ with 
$(-k,\ -k^{\prime \prime},\ -k^{\prime})$. In particular, 
note the invariance  
$qEk-QEk'\mapsto -\alpha E k+QEk''=qEk-QEk'$.
We display here the result for ${\cal{J}}_2$:\\
\footnotesize
\begin{eqnarray}
{\cal{J}}_2&=&(\frac{\sqrt{qE}}{\sqrt{\alpha E }})^{-i \epsilon
_M+1/2}(\frac{\sqrt{qE}}{\sqrt{QE}})^{i\epsilon _\mu +1/2}\Gamma
(i{\cal{E}}+\frac
12)e^{\frac{i\Omega ^2}2}e^{\frac{\pi {\cal{E}}}4}e^{-\frac{i \pi }8} D_{-i
{\cal{E}}-\frac 12} \left[
-\sqrt{2}e^{i \frac {\pi} 4}\Omega \right]\label{83}\\ 
&& (\frac{\sqrt{QE}}{\sqrt{\alpha E
}})^{2i\epsilon_M-i\epsilon_m+1/2} B(i\epsilon _\mu +\frac
12,-i\epsilon_M+\frac 12)\,_2 F_1(i
{\cal{E}}+ \frac 12,- i \epsilon _M +\frac 12,1+i(\epsilon _\mu-\epsilon
_M);\frac{qE}{QE }). \nonumber
\end{eqnarray} 
\normalsize
Therefore, the $\Br$ amplitude reads:
\footnotesize
\begin{eqnarray} 
\Br&=&
(\frac qQ)^{\frac i2(\epsilon_\mu-i\epsilon_M)}
(\frac {\alpha}Q)^{\frac i2(\epsilon_m-i\epsilon_M)}
\\ \nonumber
&&\frac {e^{-\frac \pi 2 (\epsilon_m+\epsilon_\mu)}
e^{\frac{i\pi }4}\sqrt{2\pi}
e^{-\frac i 2(\frac{k^2}{QE}\,-\frac{k^{\prime
\,2}}{qE}+\frac{k{\prime\prime\,^2}}{\alpha E })}}
{{(2QE)^{3/4}(2qE)^{1/4}(2\alpha E)^{1/4}} \Gamma (i\epsilon _m+1/2)\Gamma
(i\epsilon _\mu
+1/2)} \\ \nonumber
&&\Gamma
(i{\cal{E}}+\frac 12) e^{\frac{i\Omega ^2}2} e^{\frac{\pi
{\cal{E}}}8}e^{-\frac{i\pi }8}
D_{-i{\cal{E}}-\frac 12}\left[ \sqrt{2}e^{i\frac {5\pi} 4}\Omega \right] 
{\cal{J}}_4
\label{Bcharged}
\end{eqnarray}
\normalsize
with 
\footnotesize 
\begin{eqnarray} 
{\cal{J}}_4&=&\left\{ {e^{-\frac{3i\pi}8}}B(i\epsilon
_\mu +\frac 12,-i\epsilon_M+\frac 12)\/_2F_1(i{\cal{E}} +\frac
12,\frac 12-i\epsilon _M ,1+i\epsilon _\mu-i\epsilon
_M;\frac{qE}{Q E }) \right.  \nonumber \\
 & & +\left. {e^{-{\pi} 
\epsilon_M}e^{\frac{i\pi}8}}B(i\epsilon _m +\frac 12,-i\epsilon_M+\frac 12)
\/ _2F_1(i{\cal{E}}
+\frac 12,\frac 12-i\epsilon _M ,1+i\epsilon _m-i\epsilon
_M;\frac{\alpha E }{QE})\right\}\quad. \label{J4}
\end{eqnarray} 
\normalsize
Finally, notice that the $\Cr$ amplitude 
can be easily obtained from $\Br$ by the
substitution $m,q,k^{\prime} \mapsto \mu,
\alpha,k^{{\prime}{\prime}}$, and hence $\Omega \mapsto -\Omega$. 
To complete the calculation we have to evaluate the integrals 
${\cal{I}}_2, {\cal{J}}_2$ and ${\cal{J}}_3$. As an example we 
consider the integral ${\cal{I}}_2$, the others follow a similar pattern :  
\begin{eqnarray}
{\cal{I}}_2=\int_0^\infty dvdw\ e^{-\frac i{2QE}(\sqrt{\alpha E
}v-\sqrt{qE}w)^2+i\sqrt{2}\frac{- qE\,k+QE\,k^{\prime }}{\sqrt{QEqE\alpha E
}}(\frac{\sqrt{\alpha E }}
{\sqrt{QE}}v-\frac{\sqrt{qE}}{\sqrt{QE}}w)}\\ \nonumber
(\frac{\sqrt{\alpha E
}}{\sqrt{QE}}w+\frac{\sqrt{qE}}{\sqrt{QE}}v)^{-i\epsilon
_M-1/2}v^{i\epsilon _m-1/2}w^{i\epsilon _\mu -1/2} \label{Iresult}.
\end{eqnarray}
By using the reduced variables:
\begin{eqnarray}
X=\sqrt{\frac{\alpha E}{QE}}v\, ,Y=\sqrt{\frac{q E }{QE}}w\, ,\Omega
=\frac{k\, qE-QE k^{\prime }}{\sqrt{QEqE\alpha E }}\, ,c=e^\gamma
=\sqrt{\frac{qE}{\alpha E }}.
\end{eqnarray}
this integral becomes:
\begin{eqnarray}
\left( \sqrt{\frac{QE}{q E}}\right) ^{i\epsilon _\mu +1/2}\left(
\sqrt{\frac{QE}{\alpha E }}\right) ^{i\epsilon _m+1/2}{\cal{I}}
\end{eqnarray}
where:
\begin{eqnarray}
{\cal{I}}=\int_0^\infty dXdY\ e^{-\frac i2(X-Y)^2-i\sqrt{2}\Omega
(X-Y)}(cX+c^{-1}Y)^{-i\epsilon _M-1/2}X^{i\epsilon _m-1/2}Y^{i\epsilon _\mu
-1/2}.
\end{eqnarray}
The new changes of variables:
\begin{eqnarray}
X=\frac {\rho e^\theta}2,\ Y=\frac {\rho e^{-\theta}}2 \qquad 0<\rho
<\infty ,-\infty<\theta<\infty
\end{eqnarray}   
and $z=\rho \sinh \theta$ factorizes the double integral:
\begin{eqnarray}
{\cal{I}}&=&2^{-i {\cal{E}}+\frac 12}
\int_0^\infty dz\, d\theta e^{-\frac i2z^2}z^{ i{\cal{E}}-\frac 12} 
(\sinh \theta)^{-i{\cal{E}}-\frac 12} \\ \nonumber
&&\left\{ e^{i\theta(\epsilon_m-\epsilon_\mu)} 
(e^{\theta+\gamma}+e^{-{\theta+\gamma}})^{-i\epsilon_M-\frac 12} 
e^{-i \sqrt 2 \Omega z} \right. \nonumber \\
&+&\left. e^{-i\theta(\epsilon_m-\epsilon_\mu)} 
(e^{\gamma-\theta}+e^{-\theta+\gamma})^{-i\epsilon_M-\frac 12} 
e^{i \sqrt 2 \Omega z}\right\}
\end{eqnarray}
where the two terms come from the 
separation of the positive and negative values
 of the integration variable $\theta$. 
The $z$ integrals are the representation (\ref{DIrep}) of the parabolic cylinder
function, so:
\begin{eqnarray}
{\cal{I}}&=&2\Gamma (i\frac {\cal{E}}2+\frac 12)e^{i\frac{\Omega
^2}2}e^{\pi \frac {\cal{E}}4} e^{-i\pi/8} \int_0^\infty d\theta (2 \sinh
\theta)^{-i{\cal{E}}-\frac 12}\\ \nonumber
&&\left\{ D_{-i{\cal{E}}-\frac 12}(\sqrt2 e^{i\pi
/4}\Omega)e^{i\theta(\epsilon_m-\epsilon_\mu)}
(e^{\theta+\gamma}+e^{-{\theta+\gamma}})^{-i\epsilon_M-\frac 12} +
\right.\\  \nonumber
&&\left. D_{-i{\cal{E}}-\frac 12}(\sqrt2 e^{i\pi /4}\Omega)
e^{-i\theta(\epsilon_m-\epsilon_\mu)}
(e^{\gamma-\theta}+e^{\theta-\gamma})^{-i\epsilon_M-\frac 12} \right\}
\end{eqnarray}
A final change of variable $s=e^{2\theta}$ reduces 
the last integrals to products of Euler and
hypergeometric functions (\cite{GR}, formula 3.197.2 ). 
The final result for ${\cal{I}}_2$ reads:
\footnotesize
\begin{eqnarray}
&&{\cal{I}}_2=\left( \sqrt{\frac{QE}{q E}}\right) ^{i\epsilon _\mu
+1/2}\left( \sqrt{\frac{QE}{\alpha E }}\right) ^{i\epsilon _m+1/2}
\Gamma (\frac{i{\cal{E}}}2+\frac 12)e^{\frac{i\Omega ^2}2}e^{\frac{\pi
{\cal{E}}}4}e^{-\frac{i \pi }8} \\ \nonumber 
&&\left\{ D_{-i {\cal{E}}-\frac 12} \left[ \sqrt{2}e^{i \frac {\pi}
4}\Omega \right]B(i\epsilon _\mu +\frac 12,-i{\cal{E}}+\frac 12)\
{_2 F_1}(\frac 12 +i\epsilon_M,
i \epsilon _\mu +\frac 12,1-i\epsilon _m+i \epsilon _M;-\frac{\alpha E}{q E
}) \right. \\ \nonumber
&&\left.+D_{-i  {\cal{E}}-\frac 12} \left[- \sqrt{2}
e^{i \frac {\pi} 4}\Omega \right]B(i\epsilon _m +\frac 12,-i{\cal{E}}+\frac 12)\
{_2 F_1}(\frac 12 +i\epsilon_M,
i \epsilon _m +\frac 12,1-i\epsilon _\mu+i \epsilon
_M;-\frac{qE}{\alpha E})
\right\}
\end{eqnarray} 
\normalsize
Similarly, for the amplitude $\Br$ the last integration can be carried
out by using (\cite{GR}, formula 3.197.1).


\section{ Appendix 2: Small charge limit of the amplitudes}

In this appendix we compute the small charge limit of the amplitudes $\Ar$
and $\Br$. 
We also prove eq. (\ref{relatedS}), i.e.  that the ratio of
the amplitudes $\Ar$ and $\Br$ is given by the ratios of the Schwinger
amplitudes, up to terms proportional to  $e^{-\pi \epsilon_\mu}$.
 Moreover we shall evaluate $\Ar$ in the limit $\alpha \to 0$. As emphasize
in the main text,
a consequence of this calculation is that while the amplitudes $\Ar$ and
$\Br$ differ polynomially in the
variable $\alpha/\mu$ from their chargeless limits, their ratio is
nevertheless given by
eq.(\ref{relatedS}), which differs from the chargeless limit only by
non--perturbative
corrections. 
When $\alpha$ is small, the two
hypergeometric functions giving ${\cal J}_4$ in eq (\ref{J4}) can be
combined into one
thanks to the formulae  (9.131.2 and 9.131.1)of ref.\cite{GR}:

\begin{eqnarray}
{\cal{J}}_4&=&
\frac{e^{i\pi/8}e^{-\pi \epsilon_M}\sqrt{\alpha E}}
{Q E}
(\sqrt{\alpha E})^{-i\epsilon_M-i\epsilon_m}
(\sqrt{Q E})^{-i\epsilon_m-3i\epsilon_\mu-3i\epsilon_m}
B(\frac 12 -i\epsilon_M,\frac 12 +i\epsilon_m)\nonumber\\
&&\left\{
\frac{\Gamma(1+i\epsilon_m-i\epsilon_M)\Gamma(i\epsilon_\mu-i\epsilon_M)}
{\Gamma(\frac 12 +i{\cal{E}})\Gamma(\frac 12 -i\epsilon_M)} 
 {_2F_1}(\frac 12+i\epsilon_m, \frac 12
+i\epsilon_M,1-i\epsilon_\mu+i\epsilon_M, 
-\frac {qE}{\alpha E})\right.\nonumber \\
& & +\left. e^{-\pi \epsilon_\mu}O(\alpha/\mu) \right\}
\label{Jimp}
\end{eqnarray}
Note that (and this constitutes a check of the exactness of the
calculation), that  in the limit $\alpha/\mu \to 0$, 
there appears diverging phases in the first factor of this expression which
cancel each other.
If we omit the small corrections proportional to the Schwinger factor
$e^{-\pi \epsilon_\mu}$, most of the
prefactors are common between the amplitudes 
${\cal{A}}(k \vert k^{\prime} k^{\prime \prime})$ 
and ${\cal{B}}(k^{\prime }\vert k -k^{{\prime }{\prime }})$
and  thus can be omitted in the ratio ${\cal{A}}/{\cal{B}}$ which 
reduces to:
\begin{eqnarray}
\frac {{\cal{A}}(k \vert k^{\prime} k^{\prime \prime})} 
{{\cal{B}}(k^{\prime }\vert k -k^{{\prime }{\prime }})}&=&
e^{-i\pi/2}\frac{\Gamma(\frac 12 -i {\cal{E}})
\Gamma(\frac 12 +i {\cal{E}})e^{\pi \epsilon_M}}
{\Gamma(i\epsilon_\mu-i\epsilon_M)\Gamma(1-i\epsilon_\mu+i\epsilon_M)}
+e^{-\pi\epsilon_\mu}O(\alpha/\mu) \nonumber\\
&=&e^{-\pi(\epsilon_n-\epsilon_M)}+e^{-\pi \epsilon_\mu}O(\alpha/\mu)
\label{Jess}\end{eqnarray}   
which is the sough for result.
 Now we discuss in more details the limit of 
the amplitude $\Ar$ when $\alpha \to 0$,
and check that we recover the $\alpha =0$ result in the limit.
The computation is done in three steps. Eqs (\ref{Acharged},\ref{I3}) express
$\Ar$ as products of phases and $\Gamma$ functions with a sum of products of
Eulerian $(B)$ functions, Whittaker's ($\cal D$) functions and
hypergeometric functions.


\noindent Firstly we obtain, by a saddle point
evaluation, appropriate approximations of the Whittaker's functions. 
Then we estimate the limits of the hypergeometric functions as confluent
hypergeometric
functions. Finally, applying several times the Stirling and reflection
formulae on the
$\Gamma$-- and $B$--Euler function we obtain the result. 

\noindent The integrals 
\begin{eqnarray}
&&\Gamma (\frac{i{\cal{E}}}2+\frac 12) e^{\frac{i\Omega ^2}2}
e^{\frac{\pi {\cal{E}}}8}e^{-\frac{i\pi }8}
D_{-i\frac {\cal{E}}2-\frac 12}\left[\pm \sqrt{2}e^{i\frac {\pi} 4}
\Omega \right]=\int_0^\infty dv\ e^{\pm i\sqrt{2}
\Omega v-i\frac{v^2}2}v^{i{\cal{E}} -\frac 12}.\label{combi}
\end{eqnarray}
have saddle points located respectively at:
\begin{eqnarray}
v_s=\frac{\pm\sqrt{2}\Omega + \sqrt{2}(\Omega ^2+{\cal{E}})^{1/2}}2\qquad .
\end{eqnarray}
They are approximated by  
\be
\frac{\sqrt{2\pi }}{\sqrt{2\omega }}
e^{-\frac{i\pi }4}(2\alpha E )^{1/4}
e^{i\varphi_s^{\pm}}
\ee
 with
\footnotesize
\begin{eqnarray}
\varphi_s^{\pm}=\frac 34\frac{(k+k^{\prime})^2}{\alpha E }-\frac{\omega ^2}
{4\alpha E }\mp
 \frac{(k+k^{\prime})\omega }{2\alpha E }+\frac {{\cal{E}}}{2}
\ln \frac{\omega \mp
(k+k^{\prime})}{\sqrt{2\alpha E }}-\frac{(k+k^{\prime })(k-k^{\prime})}{2QE}\pm
\frac{(k-k^{\prime })\omega }{2QE}. 
\end{eqnarray} 
\normalsize
Let us emphasize that both integrals (\ref{combi}) have their modulus of
the same order 
of magnitude in
$\alpha /\mu$.\\
To obtain the limits of the hypergeometric functions appearing in
eq.(\ref{Ikk}) as
confluent hypergeometric functions when  $\alpha \rightarrow 0$ is
straightforward. For
small value of $\alpha$ we obtain~:
\footnotesize
\begin{eqnarray}
{_2F_1}(i\epsilon _M+\frac 12,i\epsilon _\mu +\frac 12,1+i\epsilon
_M-i\epsilon _m;-\frac{\alpha E }{qE})= \ M(i\epsilon _M+\frac
12,1+i\epsilon _M-i\epsilon _m,-\frac{i\mu ^2}{2qE})+O(\frac \alpha \mu)
\label{lim1F1},
\end{eqnarray}
and:
\begin{eqnarray}
{_2F_1}(i\epsilon _M+\frac 12,i\epsilon _m+\frac 12,1+i\epsilon
_M-i\epsilon _\mu ;-\frac{qE}{\alpha E })&=&e^{-i\frac{\mu
^2}{4qE}}(\frac{\mu ^2}{2qE})^{i\frac{\epsilon _M+\epsilon _m}2}
e^{\frac \pi 4(\epsilon _M+\epsilon _m)} \\ \nonumber
&&W_{-\frac i2(\epsilon _m+\epsilon _M),\frac i2(\epsilon _m-\epsilon
_M)}\left[ ^{-\frac{i\mu ^2}{4qE}}\right] +O(\frac \alpha \mu)
\label{limW}.
\end{eqnarray}
\normalsize
Here also these two functions are of the same order of magnitude in
$\alpha/\mu$ but the
prefactors multiplying them in the amplitude $\Ar$ are quite different. 
The first one is
multiplied by:
\footnotesize
\begin{eqnarray}
B(i\epsilon _\mu +\frac 12,-i\epsilon _\mu +i(\epsilon _M-\epsilon
_m)+\frac 12)= 
\frac{2\pi e^{i(\epsilon _M-\epsilon _m)}
e^{\pi \epsilon_M /2}e^{-\pi \epsilon_m /2}e^{-\pi \epsilon _\mu }}
{\Gamma (1+i(\epsilon
_M-\epsilon _m))} \left(1+O(\frac \alpha \mu)\right) \label{coeffA1},
\end{eqnarray} 
\normalsize
whereas the second one is pondered by:
\footnotesize
\begin{eqnarray}
B(i\epsilon _m+\frac 12,-i\epsilon _\mu +i(\epsilon _M-\epsilon _m)+
\frac 12)= \Gamma (i\epsilon _m+\frac 12)
\epsilon _\mu ^{-i\epsilon _m-1/2}e^{-\pi \epsilon_m /2}e^{i\pi /4}
(1+O(\frac \alpha \mu)) \label{coeffA2}.
\end{eqnarray} 
\normalsize
We see that the exponential factor $e^{-\pi \epsilon_\mu}$ appearing in
eq.(\ref{coeffA1}) makes
the first term of ${\cal{I}}_3$ in eq.(\ref{I3}) negligible with respect
to the second one in
the limit $\alpha \rightarrow 0$ at fixed non vanishing value of $\mu$. 
Collecting all the
results (\ref{lim1F1}, \ref{limW}) and (\ref{Jimp}), we obtain  the
limit we are discussing.
At zero order in $\alpha /\mu$, once more (as expected)  all the diverging
phases cancel in the
first term and the remaining factors group together to give the expression
(\ref{Aneutral}) with
$qE=QE$.\\ 
So, we obtain at the end: \begin{eqnarray}
{\cal{A}}(k \vert k^{\prime} k^{\prime \prime})=({\cal{A}}_0(k^{\prime}
\vert k k^{\prime \prime})+O(\alpha/\mu))+e^{-\pi
\epsilon_\mu}O(\alpha/\mu) \label{limamplA}
\end{eqnarray}
where ${\cal{A}}_0(k \vert k^{\prime} k^{\prime \prime})$ is the transition
amplitude in the neutral case (\ref{Aneutral}).
Using eq. (\ref{J4}), one can similarly show
that
\begin{eqnarray}
{\cal{B}}(k \vert k^{\prime} k^{\prime \prime})=
({\cal{B}}_0(k^{\prime} \vert k k^{\prime \prime})+O(\alpha/\mu))+
e^{-\pi \epsilon_\mu}O(\alpha/\mu) \label{limamplB}
\end{eqnarray}
where ${\cal{B}}_0 (k \vert k^{\prime} k^{\prime \prime})$ 
is the transition amplitude in the neutral case. 
from these relations (\ref{limamplA}, \ref{limamplB}) one can only a
priori deduces that their
ratio behaves as:
\begin{eqnarray}
\frac {{\cal{A}}(k \vert k^{\prime} k^{\prime \prime})} 
{{\cal{B}}(k^{\prime }\vert k -k^{{\prime }{\prime }})}&=&
e^{-\pi(\epsilon_n-\epsilon_M)}(1+O(\frac \alpha \mu))+e^{-\pi
\epsilon_\mu}O(\alpha/\mu) \quad.
\end{eqnarray}   
while, our previous computation, eq. (\ref{Jess}), shows that actually all
the polynomial
corrections in $\alpha /\mu$ to the boltzmannian factor cancel each others!

\end{document}